\documentclass[a4paper,11pt]{article}%
\usepackage{amsmath}%
\usepackage{amsfonts}%
\usepackage{amssymb}%
\usepackage{graphicx}
\usepackage{epsfig}
\usepackage{epstopdf}

\begin{document}

\title{A Study of the Electric Response of He II at the Excitation of Second Sound Waves}
\author{Tymofiy V. Chagovets
\\Institute of Physics ASCR, Na Slovance 2, 182 21 Prague,\\
Czech Republic \\} 
\date{\today}
\maketitle

\begin{abstract}
We report an experimental investigation of the electric response of
superfluid helium. Our results confirm the presence of electric
potential that appears at the relative oscillatory motion of normal fluid and
superfluid components in helium generated by the heater. The resonance
of the electric potential was observed in the first four harmonics. A
suggested method for the detection of the electric response allows the
required resonance peak to be distinguished from spurious signals.
Our results are in qualitative agreement with the data published by
previous researchers. The reasons for the discrepancy in the measured
values of the potential difference are discussed.
\end{abstract}

\section{Introduction}
\label{intro}

Superfluid helium or He II is a macroscopic quantum fluid that
exhibits extraordinary properties.
The behavior of the fluid can be understood using a Landau - Tisza model,
where He II is considered to be a two-component fluid with independent
velocity fields: the inviscid superfluid of density $\rho_s$, and the
normal fluid of density $\rho_n$, where the total density $\rho$ =
$\rho_s$ + $\rho_n$ \cite{Tisza,Landau}. The superfluid has neither
viscosity nor entropy, and the entire heat content of He II is carried
by the normal component. This simplified picture is described by two
fluid equations of motion. 
One of the important outcome of these equations is not only the existence of density fluctuations relative to ordinary or first sound, but also a prediction of the second sound, a wave described by temperature fluctuations.  

Recent experiments \cite{Rybalko_LTP,Rybalko_JLTP} reported another exceptional property of superfluid $^4$He, such as electrical activity that appears at the relative oscillatory motion of the normal fluid and superfluid components, or second sound wave.
A standing half-wave of second sound was generated by a heater, which was placed on one end of a dielectric resonator. 
The AC potential difference between an electrode that was placed on the opposite end of the resonator and the ground was recorded. 
The resonance frequency of the electric response corresponded to the frequency of the second sound resonance.
The experiments showed that the amplitude of the electric potential,
$\Delta U$, is proportional to the amplitude of the temperature oscillations, $\Delta T$, in a second sound wave.
The ratio $\Delta T / \Delta U \approx 2e / k_{B}$ was independent of the resonator size and the temperature of the bath in the range 1.4-1.8 K.
Here, $e$ is the electron charge, and $k_{B}$ is the Boltzmann constant. 
It should be noted that experiments with the first sound in both He I
and He II did not show any electrical signal, even when high power was applied to the sound emitter \cite{Rybalko_LTP}.
This electrical activity of superfluid helium is rather puzzling. The
helium atom is not only chemically inert and electrically neutral and does not have an electric dipole moment, but it also has the highest ionization potential of any element and, correspondingly, an exceptionally low polarizability.

The experiments of Ref. \cite{Rybalko_LTP,Rybalko_JLTP} stimulated a
large number of theoretical studies and suggested various theoretical
explanations for this phenomenon \cite{Natsik_3,Kosevich_1,Melnikovsky,Tomchenko_1,Tomchenko_2,Tomchenko_3,Shevchenko_1,Pashitskii}. 
For instance, in Ref. \cite{Kosevich_1}, it was assumed that superfluid helium has an ordered electric quadrupole moment that can be polarized by the nonuniform flow of the superfluid. However, the microscopic nature of such a quadrupole moment was not discussed and remains an unanswered question.
The author of Ref. \cite{Melnikovsky} suggested the inertial mechanism of polarization of $^4$He atoms based on the large mass difference between electrons and nuclei.
An analog of this polarization occurs in metals and is called the Stewart -- Tolman effect.
However, this approach supposes the polarization of helium by first
sound, which contradicts the experimental results of Ref. \cite{Rybalko_LTP}. 
Moreover, this theoretical model predicts a temperature dependence for
the $\Delta T / \Delta U$ ratio, which also contradicts the experimental results. 
In Ref. \cite{Natsik_3}, the electrical activity was interpreted based
on a similar inertial effect as in Ref. \cite{Melnikovsky}, which
arises under the influence of a centrifugal force causing a nonuniform
azimuthal rotational velocity of the superfluid component around the
axis of a quantum vortex in He II. However, in this case, macroscopic
polarization requires a relatively high concentration of vortex lines.
Thus, the theoretical models are discussed with regard to the possible explanations and contradictions that arise in understanding the experimental data, but a satisfactory explanation of this phenomenon has not yet been found.

Additionally, detailed analysis of the articles \cite{Rybalko_LTP,Rybalko_JLTP} revealed a number of questions that need to be addressed. 
For example, the amplitude of the signals is relatively small (approximately 100 nV), and measuring such signals can be difficult due to various sources of noise. 
In this report, we will demonstrate the existence of a large number of
spurious peaks in a frequency sweep that have amplitudes comparable to
that of the signal of interest. 
These spurious peaks can potentially be misinterpreted as the required electric response. 
In the original experiments, the authors used a measuring loop with the compensation of the input capacitance. 
Such a measuring circuit itself can be a source of electric pickup or spurious signals with amplitudes comparable to the observed electric response. 
Unfortunately, the original papers \cite{Rybalko_LTP,Rybalko_JLTP} do
not include some rather important information, such as the principal
scheme of the measuring circuit, the capacitance of the connecting
wires and electrodes used in the experiment, and the area of the resonator's cross-section. 
Moreover, the analysis of the experimental data shows some contradictions in experimental data reported in \cite{Rybalko_LTP} and \cite{Rybalko_JLTP} that reduce confidence in these results and  may give a most misleading impression. For example, $\Delta$T values, reported in \cite{Rybalko_LTP} and \cite{Rybalko_JLTP}, is supposed to be related to the same experiment, however they are approximately one order of magnitude higher. Therefore, the electric response in He II is require detailed experimental investigation with confirmation of some previous results.

The singularity of the phenomenon, the relatively small amplitude of
the measured signal, and the lack of some important experimental details in the original papers motivated further study of the electric response in He II. 
The absence of new experimental data and of an appropriate theoretical
model for the last 10 years makes this issue even more interesting. 
The purpose of this work is detailed verification of correlation between the second sound resonance and the resonance of electric response, test the existence of the electric response on higher modes of the second sound. The investigation this phenomenon is also requires a test various measuring circuits without compensation of the input capacitance and develop a reliable method of detecting the electric response signal in He II.


\section{Experimental setup}
\label{sec:1}
Our experiments were performed in an open bath cryostat, and a pumping
unit was used to maintain the temperature in the range 1.4-2.2 K. The
helium bath temperature was monitored and stabilized at the chosen
value using a Conductus LTC-21 temperature controller with a
calibrated Cernox sensor and a 50 $\Omega$ heater mounted on the
bottom of the cryostat. With this system, we were able to stabilize
the bath at temperatures of up to $\pm$ 100 $\mu$K.

The resonator, which was 25-mm long and had an inner diameter of 7 mm,
was machined from a piece of epoxy, and its inner surface was
polished. Both ends of the resonator can be covered by flanges with
various attached sensors, for example, a heater with bifilar wire or a
thin-film thermometer, or mounted with electrodes. 
The resonator was hung on a support inside a metallic shielded box and had no electrical contact with the metallic parts of the cryostat.
All sensors were
connected to the measuring circuit with low temperature coax cables
\cite{comm}, and the total capacity of the loop was approximately 260 pF.

To excite the second sound wave, we used a manganin heater with
approximately 180 $\Omega$ resistance, which was glued with Varnish glue to the
surface of a polished brass disk with a thickness of 4 mm that was
connected to one end of the resonator. The  heater, which had a spiral
shape, was made of two layers bifilar wire that covered the entire
cross-section of the channel.  

The initial experiments \cite{Rybalko_LTP,Rybalko_JLTP} showed low
peak values of the resonance of approximately $10^{-7}$ V with a peak
width of approximately 2 Hz. 
Identifying such small signals is a rather complicated task due to the
relatively high noise level, which was only 5 times lower than the magnitude of the signal. 
The low signal-to-noise ratio and small width of the resonance peak
require measurements with a large time constant (usually 1 or 3 seconds) for the frequency
sweep with a step of approximately 10 mHz. 
The increase in the number of points allows to receive a resonant curve of more regular shape.
These types of measurement occur over a long period of time, and the
identification of the resonance peak is difficult because of the wide range of frequencies. Nevertheless, the range of interest can be determined analytically. 

As was discovered in Ref \cite{Rybalko_LTP}, the resonance frequency
of the electric response coincides with the resonance frequency of the second sound. 
The resonance frequency, $f_r$, of second sound can be calculated as:

\begin{equation}
\label{eq:1}
f_r = \frac{1}{4}\frac{n u_2}{L}
\end{equation}
\noindent
where $u_2$ is value of the second sound velocity, $L$ is the actual
length of the resonator and $n$ is the number of the half-wave lengths
of the second sound signal for the observed resonance mode. The factor
1/4 is caused by doubling the emission frequency of the heater and
because the first resonance occurs when there is half of a second
sound wavelength between the transmitter and receiver.

\begin{figure}
  \includegraphics[scale=0.5]{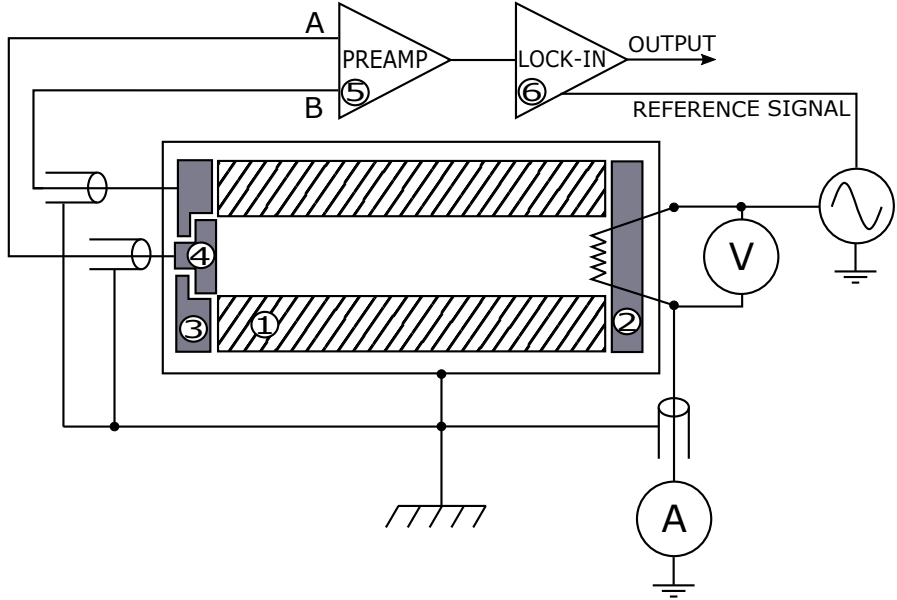}
\caption{Electrical schematic of a typical resonator: (1) epoxy body
  of the resonator, (2) brass plate with heater, (3) brass mount with
  (4) brass electrode for detecting the electric response, (5)
  low-noise voltage preamplifier SR 560 from Stanford Research
  Systems, and (6) SR 830 lock-in amplifier.}
\label{fig:1}       
\end{figure}

The actual length of the resonator was defined using a set of
measurements of the second sound resonance. For this purpose, the
resonator end opposite to the heater was covered by a polished brass
disk with a Ge film thermometer glued to the center of the disk
\cite{Mitin}. The low-inertia thermometer ensured the complete thermal
tracking of the second sound waves. The thermometer was energized with
a 1 $\mu$A DC current; to detect the signals, we used a SR830 dual phase lock-in amplifier. 
To excite the second sound wave, we used an Agilent 33250A function
generator to drive the sinusoidal pattern of the heater.

The wave forms of the resonant modes are sinusoidal in space, the
velocity fields have nodes at the ends of the resonators, and the
temperature oscillations have antinodes coinciding with the velocity
nodes. The observed second sound resonance peaks have Lorentzian
shapes. Our resonator had a quality factor of approximately 60, and
the actual length of the resonator L = 24.3 mm. The relatively low
value of the resonator's actual length, as compared with its
geometrical length, may be attributed to the heater thickness, which
is slightly larger in the resonator.
The non-flat form of the heater is presumably a reason for the low
value of the quality factor.

\section{Electric response}
\label{sec:2}

To study the electric response of He II in the resonator generated
by a standing half-wave of second sound, the plate with the
thermometer was replaced by a cylindrical electrode. The electrode,
which had a thickness of 3 mm and a diameter of 7 mm, was glued with Stycast \textit{in situ} inside the brass mount.

We considered several measuring circuits depending on the nature of
the phenomenon and the quality of the detected signal. 
Measurements of the electric potential between the plates fixed on both ends of the resonator would be a relatively reasonable way to study the bulk polarization of helium atoms.
In this case, the electrode and the brass plate with the heater had a
differential voltage connection (A-B) to the lock-in, whereby coaxial
cable (A) was connected to the electrode and coaxial cable (B)
connected to the plate on the opposite side of the resonator. The
observed noise level in this case was approximately 10$^{-5}$ V, even
for small drive amplitudes, whereas the amplitude of the required
signal was supposed to be on the order of 10$^{-7}$ V. As a result,
the high noise level made signal detection completely
impossible. Although there was no electric connection between the
heater and the plate, the proximity of the heater to the plate led to
the appearance additional electromagnetic noise between the electrodes.

\begin{figure}
  \includegraphics[scale=0.45]{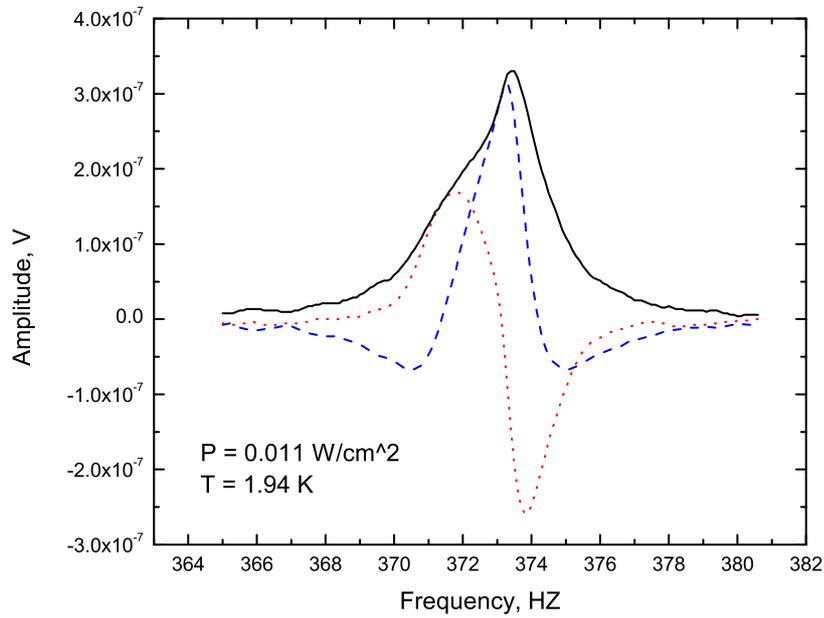}
\caption{(Color on-line) The form of the second mode of the electric
  response signal plotted against frequency, showing the absorption
  (blue dashed curve) and dispersion (red dotted curve) data measured
  at 1.94 K with AC heat flux 0.011 W/cm$^2$. The black curve
  corresponds to the signal magnitude.}
\label{fig:2}       
\end{figure}

Another possible method is to measure the potential difference between
the electrode and the ground. In this circuit, the noise of the signal
was approximately one order of magnitude lower than in the previous
case. It was possible to observe a resonance peak, but only at higher
drive amplitudes ($\sim$ 10*10$^{-3}$ W/cm$^2$). We
assume that the noise that arises from the coaxial cable's shield is
the reason for the low resolution of the observed signal.

A significant enhancement of the signal-to-noise ratio was achieved
when the electrode and its mount were connected to the lock-in
differentially. Figure \ref{fig:1} shows a principal scheme of the
resonator with the electrical connections that we used to study
electrical activity in He II. Basically, the lock-in measures the voltage
difference between the center conductor of  coaxial cable (A)
connected to the electrode and coaxial cable (B) connected to the
electrode's mount. 
Noise pickup on the shield has no influence on the noise in the signal because the shields are ignored.
The capacitance
between the electrode and the mount was approximately 20 pF (losses
0.025), and the total capacitance of the cables was $\sim$ 260 pF
(losses 0.02). In some experiments, a low-noise voltage preamplifier
with differential voltage input (SR 560 from Stanford Research
Systems) was added to the measuring loop. The use of a preamplifier
with a bandwidth filter helped to reduce the interfering noise that
appears due to mechanical vibration of the nanovolt level. This scheme
allows for the measurement of charge density only in a thin layer of
helium near the electrode. Thereby, it is difficult to determine the
type of charge distribution that takes place in the resonator volume.

\begin{figure}
  \includegraphics[scale=0.4]{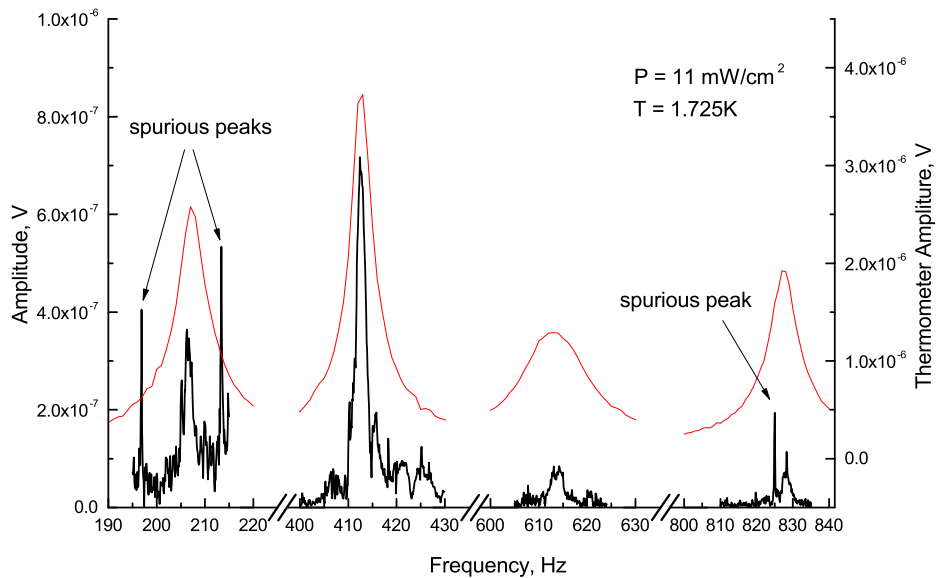}
\caption{(Color on-line) The resonance peaks of the electric response
  of the first four harmonics (bottom black curves) measured at 1.73
  K. The top red curves represent the experimental data collected for
  the second sound resonances using the same harmonics.}
\label{fig:3}       
\end{figure}

The electric response generated by the second sound wave was observed
in the temperature range 1.71-2.04 K. An example of the electric
response resonance is presented in Figure \ref{fig:2}. 
The amplitude and shape of the observed resonance peaks were very
sensitive to the temperature stabilization of the helium bath. The
magnitude of the resonance dramatically decreases when the temperature
fluctuation is on the order of 500 $\mu$K; thus, the signal was barely
distinguishable from the background noise.

The signal of the electrical activity was observed using the first
four harmonics (see Figure \ref{fig:3}). In most cases, the highest
peak amplitude was observed for the second harmonic, while for the third and fourth harmonics, the signal was almost lost in noise.
The resonance frequencies of all four harmonics are in a good
agreement with the resonance frequencies of the second sound that was independently measured in the same resonator.

It should be noted that a difference between the width of second sound resonance and the width of electric response was not observed in Ref. \cite{Rybalko_LTP}.
In our experiment, the width of second sound resonances was approximately two times higher than the width of electric response. 
Moreover, the values of width could change from measurement to measurement.
As noted in Ref. \cite{Zinoveva}, the reason for this low accuracy is low quality factor of the resonator.

In addition to the main peak that represented the electric response, we also detected a number of spurious signals.
For example, there are two sharp peaks in Figure \ref{fig:3}, which are located at the left and right sides of the first resonance mode. 
The magnitudes of these peaks are comparable or even higher than the required signal.
We assume that mechanical vibrations from the pumping unit lead to a
microphonic effect in the coaxial cables that appears as resonance peaks of a certain frequency.

\begin{figure}
  \includegraphics[scale=0.4]{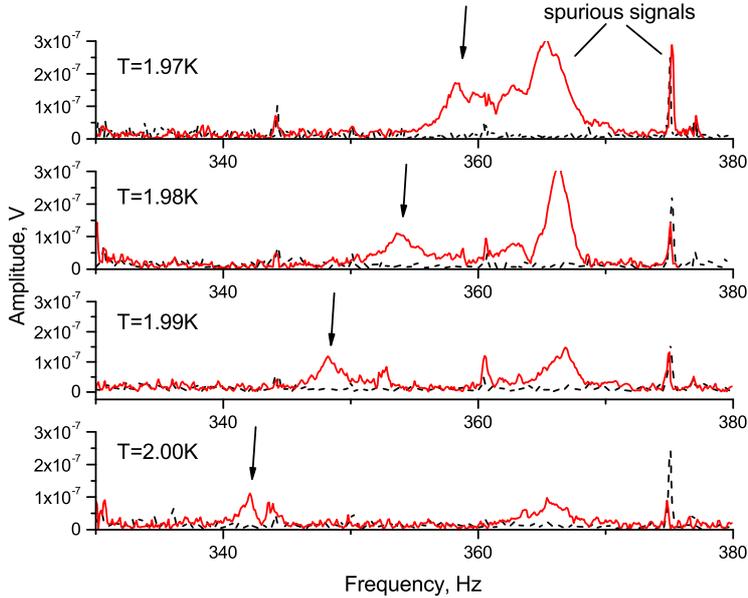}
\caption{(Color on-line) Changing the resonance frequency of the
  standing wave by changing the bath temperature. The frequency range
  corresponds to the resonance frequency of the second harmonics in
  the temperature range 1.97 - 2K. 
The AC power heat flux was approximately 0.011 W/cm$^2$ at all temperatures. 
The black dashed curves correspond to the frequency sweet with zero excitation current applied to the heater.
Black arrows indicate the frequency of electric induction resonance.
}
\label{fig:4}       
\end{figure}

For further study, we choose the second harmonic because of the higher
amplitude of the signal and smaller number of spurious peaks compared with the first mode.
Nevertheless, a frequency sweep in a range of interest chosen nearby
the frequency of the second sound resonance revealed a number of peaks
that might be misinterpreted as the signal of electrical activity in
He II (see the red curve in Figure \ref{fig:4}, top graph). 
To exclude the possible influence of electric pickup on the signal,
the frequency was swept with zero excitation current applied to the
heater (see the black dashed curve in Figure \ref{fig:4}). 
In this case, the resonance signal was not observed in the expected
frequency range, but it was also observed that some spurious peaks disappeared from the frequency sweep.

Taking into account the difficulty in defining the required signal, the following measuring protocol was implemented. 
As was shown in Ref.\cite{Rybalko_LTP,Rybalko_JLTP}, the resonance
frequency of the electric response is strongly dependent on the temperature.
Performing measurements at several temperatures that were slightly different (about 10-30mK) from the specified temperature revealed a frequency shift for one of the peak signals.
The shift of the resonance frequency was proportional to the change in
the second sound velocity at the given temperature. Thus, this peak
was defined as a resonance of the electric response.
Figure \ref{fig:4} shows the frequency sweep for 4 different temperatures. 
The resonance that represents the electric response (marked by black
arrows) shifts gradually as the temperature varies. 
There are a number of other peaks in this frequency range, but the
resonance frequencies of these peaks is temperature
independent. Therefore, we consider them to be spurious peaks.

\section{Discussion}
\label{sec:3}

\begin{figure}
  \includegraphics[scale=0.45]{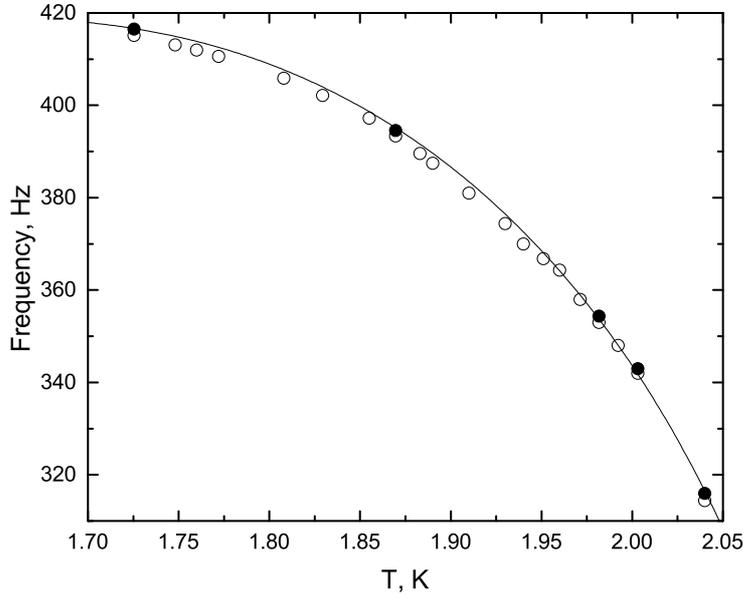}
\caption{Temperature dependence of the resonance frequency of the electric response (open circles). The closed circles correspond to the resonance frequency the second sound. The solid line represents the resonance frequency of the second sound calculated for the resonator based on known data of the second sound velocity.
}
\label{fig:5}       
\end{figure}

The dependence of the resonance frequency for the second mode as a
function of temperature is presented in Figure \ref{fig:5}.
These data can be compared with the value of the second sound resonance frequency calculated for our channel using Eq. \ref{eq:1}.
A small discrepancy between the calculated value of the resonance
frequency and the value observed in the experiment can be explained by
the slight differences in the geometry of the resonator. 
We calculated the resonator length based on the resonance frequency of
the second sound observed in our resonator. 
The thermometer used in the second sound experiment had a thickness of
approximately 1 mm and was slightly recessed into the cavity, which reduced its actual length. 
As a result the resonance frequency of the second sound is shifted
toward higher frequencies with respect to that of the electric response.
For example, decreasing the resonator length by 1 mm increases the
resonance frequency by approximately 7 Hz. In our experiment, the
difference between the experimental and calculated values is
about 4 Hz, which is consistent with the difference in length
between the two experiments.

Measurements of second sound resonance can be made simultaneously with a measurement of the electric response. 
In this case, the thermometer and electrode must be placed close to each other on one side of the resonator. 
However, the periodic oscillation of voltage on the thermometer due to the temperature oscillation in the second sound wave may induce some pickup signal on the electrode. 
To avoid potential sources of electric pickup, the measurements of electric response and second sound resonance were made in two different experiments. 

Despite the low quality factor of the resonator in comparison with previous research, the absolute values of resonance amplitude observed in our experiment are higher than the one observed previously. 
The ratio $\Delta T / \Delta U$ is on the order of one hundred, while
the value presented in the original article was two orders of magnitude higher.
We believe that the discrepancy between the two experiments can be
understood as follows.  
In both experiments, a potential difference, $\Delta U$, was measured
on the electrode. The magnitude of $\Delta U$ can be determined as
$\Delta U = \Delta Q / C_{in}$, where $C_{in}$ is the input capacitance. 
However, the parameters of the electrode, such as its capacitance and
cross-section, were different in the two experiments. Therefore, the
value of $\Delta U$ must be reconsidered in terms of the number of charges per unit area.
In our resonator, this value is on the order of 1000 for the similar
$\Delta T$ values reported in the original papers. The reports \cite{Rybalko_LTP,Rybalko_JLTP} do not provide information about the capacitance or the surface area of the electrode, which makes it difficult to compare the results. 
It can be estimated that the discussed parameters of the electrode were somewhat smaller in the original experiment.

\section{Conclusions}
\label{conclusions}

We have presented an experimental study of the electrical activity of superfluid helium that appears at the relative oscillatory motion of the normal fluid and superfluid components.
The created apparatus allowed us to register an A.C. electric potential of approximately 5*10$^{-7}$ V.
Our data give suggest the presence of an electric potential in
superfluid helium excited by a second sound standing wave. 
For the first time, the electric response was observed for the first
four harmonics with resonance frequencies that were similar to the frequency of second sound resonance. 
The resonance frequency of the electric response was compatible with
the frequency of second sound resonance in the temperature range
1.72-2.04K.
Despite the weak connection between the inner space of the resonator
and the helium bath, the amplitude of the resonance peak was very
sensitive to thermal fluctuation of the helium bath. We suggested a
reliable method for the detection of the electric response that allows
the required resonance to be distinguished from spurious signals. 
Our results are in qualitative agreement with the data published by previous researchers. 
Nevertheless, the amplitude of the resonance peak was higher despite
the lower quality factor of the resonator. The reason for this
discrepancy in the absolute values of the potential difference lies in
the parameters of the electrode, such as the cross-section and capacitance.
We believe that further investigation of this phenomenon could lead to a more complete understanding of the electric properties of superfluid helium.

\paragraph{Acknowledgements:}
The authors appreciate the technical assistance of F. Soukup and
stimulating discussions with A. Rybalko, E. Rudavskii, L. Skrbek and
V. Chagovets. This work is supported by the Czech Science Foundation (GACR) under Project No. 13-03806P.


\begin{thebibliography}{99}
%
%
\bibitem{Tisza} 
L. Tisza, Nature \textbf{141} 913 (1938)

\bibitem{Landau}
L. D. Landau, J. Phys. (USSR) \textbf{5}, 71 (1941)

\bibitem{Rybalko_LTP} A.S. Rybalko, Low Temp. Phys. \textbf{30}, 994 (2004)
\bibitem{Rybalko_JLTP} A.S. Rybalko, et al. J. Low Temp. Phys. \textbf{148}, 527 (2007)

\bibitem{Natsik_3} V. D. Natsik, Low Temp. Phys. \textbf{34}, 493 (2008)


\bibitem{Kosevich_1} A. M. Kosevich, Low Temp. Phys. \textbf{31}, 37 (2005)

\bibitem{Melnikovsky} L. A. Melnikovsky, J. Low Temp. Phys. \textbf{148}, 559 (2007)

\bibitem{Tomchenko_1} M. D. Tomchenko, Phys. Rev. B \textbf{83}, 094512 (2011)
\bibitem{Tomchenko_2} V.M. Loktev, M. D. Tomchenko, J. Phys. B At. Mol. Opt. Phys. \textbf{44}, 035006 (2011)
\bibitem{Tomchenko_3} M. D. Tomchenko, J. Low Temp. Phys. \textbf{158}, 854 (2010)

\bibitem{Shevchenko_1} S. I. Shevchenko, A. S. Rukin, Low Temp. Phys. \textbf{37}, 884 (2011)

\bibitem{Pashitskii} E. A. Pashitskii, A. A. Gurin, J. Exp. Theor. Phys. \textbf{111} 975 (2011)

\bibitem{comm} In our experiment we use LakeShore coax cable with nominal capacity about 170 pF/m.

\bibitem{Mitin} V. F. Mitin, Advances in Cryogenic Engineering \textbf{43}, (1982)

\bibitem{Zinoveva} K. N. Zinoveva, Sov. Phys. JETP \textbf{25}, 2(8), 235 (1953)

\end{thebibliography}
\end{document}